\documentclass[reprint,amssymb, amsmath, aip,cha,twocolumn]{revtex4-1}
\usepackage{graphicx}
\usepackage[caption = false]{subfig}
\usepackage{color}
\usepackage{epsfig}
\usepackage{ifpdf}
\usepackage{url}%
\usepackage{bm}
\usepackage[colorlinks=true,linkcolor=blue]{hyperref}%
\usepackage{cleveref}
\expandafter\ifx\csname package@font\endcsname\relax\else
 \expandafter\expandafter
 \expandafter\usepackage
 \expandafter\expandafter
 \expandafter{\csname package@font\endcsname}%
\fi
\begin{document}
\title{Anomalous time of flight behavior of fast ions in laser produced aluminum plasma}%
\author{Garima Arora}%
\email{garimagarora@gmail.com}
\affiliation{Institute For Plasma Research, Bhat, Gandhinagar,Gujarat, 382428, India}
\author{Jinto Thomas}
\affiliation{Institute For Plasma Research, Bhat, Gandhinagar,Gujarat, 382428, India}
\author{Hem Chandra Joshi}
\affiliation{Institute For Plasma Research, Bhat, Gandhinagar,Gujarat, 382428, India}%
\affiliation{Homi Bhabha National Institute, Training School Complex, Anushaktinagar, Mumbai, 400094, India}
\date{\today}
\begin{abstract} 
 In this work, dynamics of multi-charged ions emitted from an aluminum plasma produced by Q switched Nd: Yag laser is studied using time of flight (TOF) measurements from Langmuir Probe (LP) and spectroscopy (STOF) under Ar ambient of 0.02 mbar. The temporal evolution of multi-charged ions, background neutrals and ions is systematically studied for varying laser intensities. The temporal evolution shows all the species have double peak structure for all the laser intensities considered in the study. The fast peak is sharp whereas the slow peak is broad similar to that observed in previous studies. Moreover, higher charged ions have higher velocity, indicating acceleration from the transient electric field produced at the very initial temporal stages of expansion. Interestingly, the fast peak gets delayed, whereas the slow peak advances in time with increased laser intensity, which has not been reported in earlier studies. The observations point towards the possible role of ambipolar electric fields in the unexpected observed behavior of the TOF profiles.
\end{abstract}
\maketitle
\section{Introduction}\label{sec:intro}
 Laser produced plasma has been studied extensively because of various applications such as intertial confinement fusion \cite{hora2007new}, laser induced breakdown spectroscopy (LIBS) \cite{loudyi2009improving,noll2018libs,article_garima}, lithograpghy \cite{stamm2004extreme}, development of ion source \cite{balki2018optical}, extreme ultravoilet light source \cite{stamm2004extreme} pulsed laser decomposition (PLD)\cite{geohegan1994pulsed}, plasma diagnostics \cite{harilalrecent,wu2017diagnostic,harilal1997electron,verhoff2012angular}, nanoparticle generation \cite{rao1995nanoparticle}, etc. Laser ablation is a dynamic process in which the removed material is converted into vapor or partially ionized cloud when the laser strikes the surface of the target. The plasma plume, thus formed, is composed of electrons of few eV and energetic ions of hundreds and thousands of eV. \par 
The energy distribution of ions far away from the target can be divided into two groups; (1) low energy ions comprising of ablated mass and (2) a smaller fraction of energetic, fast ions. These fast ions from laser produced plasma form a significantly small group of ions that are the transporters of absorbed laser energy \cite{elsied2016characteristics}. The study of the dynamics of fast ions from laser produced plasma has significant interest for inertial confinement fusion \cite{hora2007new}, laser-matter interaction\cite{tajima2002zettawatt,daido2012review} and different accelerations mechanisms \cite{amiranoff1998observation,sprangle1988laser}.  Hence, a systematic study of ion acceleration is naturally quite informative in furthering the understanding of laser-matter interaction process. Spectroscopic time of flight (STOF) 
appears to be an important diagnostics for understanding
the laser-plasma interaction \cite{harilal2022optical,ma14237336}.\par
Acceleration of ions with ns or ultrafast laser pulses due to ambipolar electric field originating from double layer (DL) structures has already been reported \cite{bulgakova2000double,thomas2020observation}. Bulgakawa \textit{et. al.} \cite{bulgakova2000double} explained the origin of the double-peak structure in ion collector time of flight distribution (TOF) in case of vacuum as well as in the presence of background gas due to  self-generated ambipolar electric field or fromation of DL. DL is formed by the spatial charge separation in the expanding plasma and consequently results in disrupting the quasineutrality condition. Thus, DL modulates the velocity distribution of ions into fast and thermal \cite{bulgakova2000double,wu2020dynamic,batool2021time} and forms double peak structure. Wu \textit{et. al.} \cite{wu2020dynamic} studied the dynamics of multi-charged ions emitted from ns laser produced molybdenum plasma from mass spectrometry. They observed multi peak structures in time-resolved mass spectroscopy and linked the velocities of the multi-charged ions to the acceleration in the transient sheath. Further, they have reported saturation of kinetic energy of molybdenum ions with laser intensity.
Batool \textit{et. al.}\cite{batool2021time} observed two distant peaks of ions with time delay of nanoseconds and microseconds due to charge separation between the prompt electrons and initially emitted ions. Dogar \textit{et. al.}\cite{dogar2017characterization} measured ion flux from TOF ion collector for various metals using ns laser pulse. They have observed two  groups of fast ions. Further, they reported that early fast peak is present in case of all the metals used in the study and attributed it to surface contamination. The second group is suggested to be due to ion acceleration from the prompt electron emission and is observed only in case of heavier metals. Bhatti \textit{et. al.} \cite{bhatti2021energy} suggested that self generated electric field (SGEF) should be responsible for the fast ions. They observed increase in SGEF with increase in laser irradiance. Farid \textit{et. al.} \cite{farid2013kinetics} reported enhancement in kinetic energy for both fast and slow ions with increase in laser fluence which is saturated at higher fluence.\par
Despite these report works, concerted study of the time evolution of various charge states appears to be quite challenging for getting better understanding of the involved processes. Hence, in this work we have systematically studied the effect of laser fluence  on the charge states and background argon in laser produced aluminum plasma with Langmuir probe (LP) and spectral time of flight (STOF) diagnostics at a background pressure of 0.02 mbar argon ambient. In line with earlier works, double peak structure is observed in the temporal evolution from both the diagnostics. Multi-charged ions exhibiting dependence on arrival time with laser fluence are also observed. Interestingly in contrast to the anticipated behavior, the fast ions appear to get delayed with laser intensity. This peculiar behavior is attributed to possible decrease in SGEF.\\
  The paper is divided in  the following sections. Sec II gives the details of the experimental setup, laser and plasma diagnostics used to investigate dynamics of fast ions. Sec III  describes the observations and discussion. Concluding remarks are in Sec. IV.

\section{Experimental Set-up}\label{sec:setup}
 \begin{figure}[ht]
\includegraphics[scale=0.4]{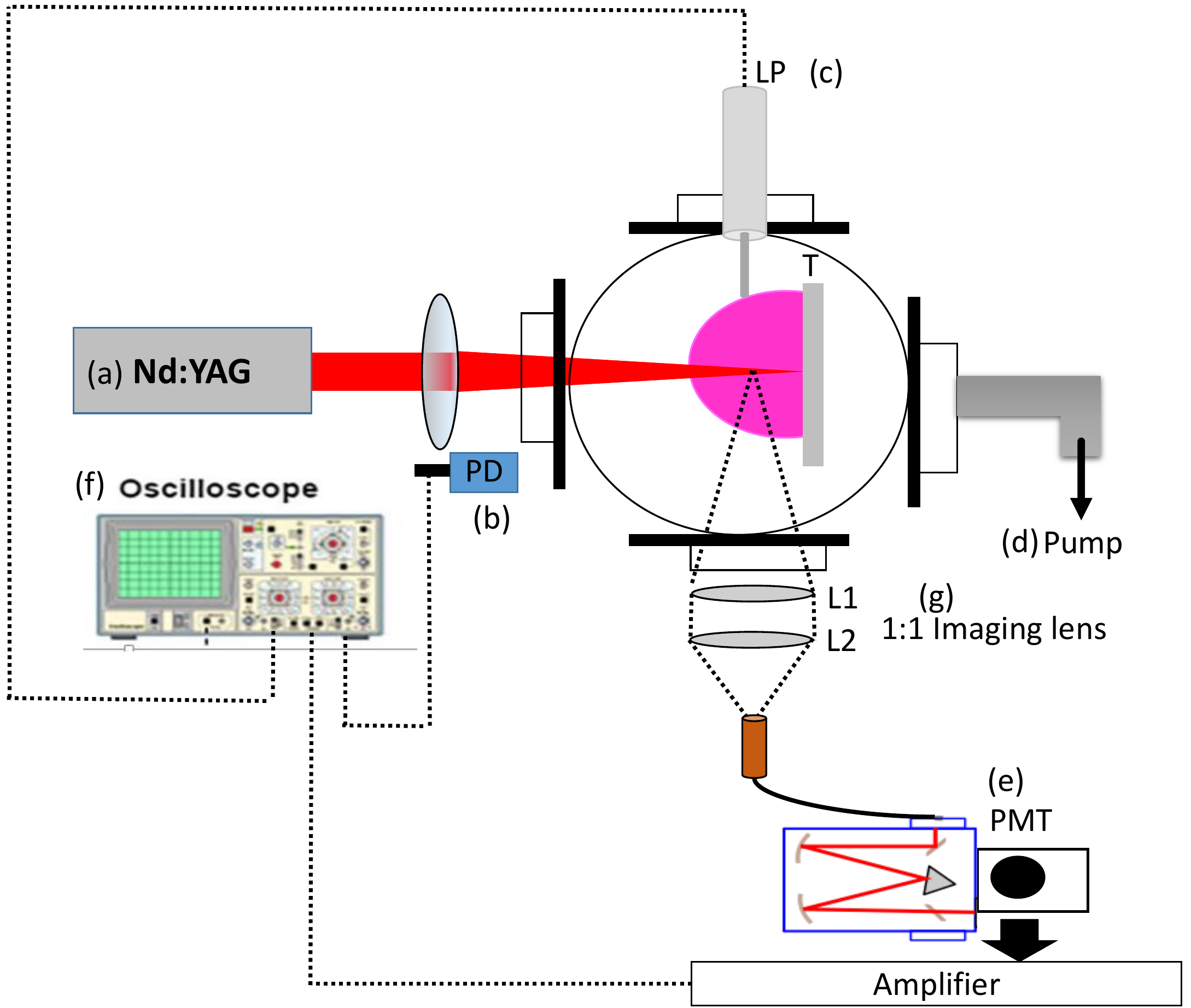}
\caption{\label{fig:fig1} Schematic diagram of experimental set-up. (a) ND:YAG laser,(b) PD photodiode (c) LP single Langmuir Probe (LP) (d) Rotary Pump (e) Phtot Multiplier Tube (PMT) (f) Fast Digital Oscilloscope and (g) L1 and L2 lens system. }
\end{figure}
Fig.\ref{fig:fig1} shows the schematic diagram of the experimental set-up used to study the dynamics of ions using LP and STOF. A Q switched Nd:YAG laser (wavelength $\lambda=1064 nm$, pulse width $\tau=8 ns$) is used to ablate the target and produce plasma. The energy of the laser is varied from 100 mJ to 930 mJ by changing the delay between the flash lamps of the oscillator and amplifier unit of the laser. The laser is focused on the sample using a 25 cm plano convex lens and the lens position is adjusted to get a spot size of $\sim 1 mm$ on the sample resulting in laser intensity in the range of 3-24 $GW/cm^2$.  A fast photodiode (PD) is placed near the focusing lens to pick part of the reflected laser light to record the time of incidence of the laser on the sample.  The PD triggers the fast digital oscilloscope, and ICCD to record the signal. \par
A cylindrical glass chamber of 100 mm diameter and 600 mm length evacuated by a rotary pump is used for the experiments. The chamber is filled with argon gas to set the desired working pressure. A pirani gauge is used for measuring the pressure in the chamber. Due to experimental limitations of the present chamber, the working pressure is set to 0.02 mbar. A well polished and cleaned aluminum strip  mounted on a vacuum compatible translation stage is used as the target for laser produced plasma. The translation stage is moved after each laser pulse so that fresh sample location is used for each experiment. Multiple acquisitions are performed for each experiment to reduce the statistical error. Various diagnostics such as LP, STOF, 1 m long Czerny-Turner spectrograph coupled with ICCD, and fast imaging with different narrow-band interference filters are used for studying the fast and slow dynamics of the plasma plume. \par
 LP is made of tungsten wire having a diameter of 1 mm and length 10 mm is mounted on a linear motion feed through to position it accurately within the plasma plume. It is placed orthogonal to the plasma plume propagation and its tip is placed 2-3 mm away from the laser beam path to avoid obstruction to the laser path. LP measurements are performed at 30 mm from the sample and the probe can be considered almost 90 degree to the plasma plume propagation.  The probe is negatively biased to -30 V in series with a 50 $\Omega$ resistance, and the current is measured using a high resolution (5V/A) current transformer.\\
 An imaging system consisting of two lenses and magnification 0.5 is used to collect the emission from the plasma plume into an optical fiber. The optical fiber is of 600 microns core diameter and provides a spatial resolution of 1.2 mm for the STOF measurements. The fiber is coupled to a monochromator (Hr 460) with slit width $\sim$ 50 microns. A high gain fast photomultiplier tube  (PMT) is used as the detector. The PMT output is connected to a buffer amplifier and fed to a fast digital oscilloscope to record the temporal evolution. Another 1 m long Czerny-Turner spectrograph coupled with ICCD (Andor DH734) is used to record the emission spectra from the plasma plume. The LP and STOF data are averaged over five plasma shots to reduce the statistical uncertainty in the measurements. 
\section{Results and Discussion}\label{sec:results}
\subsection{Time of Flight Measurements}
 \begin{figure}[ht]
\includegraphics[scale=1.0]{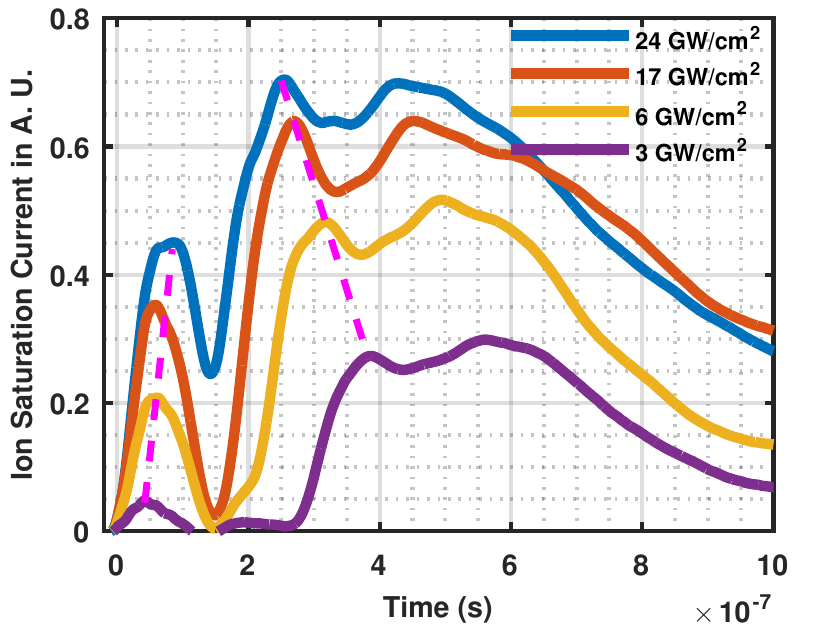}
\caption{\label{fig:fig2} Temporal evolution of ion saturation current of plasma plume for four different laser intensities recorded at 30 mm from the target. The plasma plume expands into argon background pressure of 0.02 mbar. The dotted pink lines are formed by joining the peaking time of ion current profile to mark the evolution of peaking times of fast and slow ions.   }
\end{figure}
Figure $\ref{fig:fig2}$ shows the temporal evolution of ion saturation current measured using the LP at a distance of 30 mm from the sample at an ambient pressure of 0.02 mbar for different laser intensities. LP is kept normal to the plume expansion as shown in the figure. As the validity of LP measurements depends upon the plasma plume conditions \cite{kumar2009parametric}, care has to be taken while making the measurements. However, in this work the LP simply used for studying the temporal evolution of the ions. As can be seen in the figure, the ion saturation current has two prominent peaks. The first peak observed at a delay of 100 ns is due to the fast ions generated from the laser plasma. The fast ions can be attributed to accelerated ions from the electric field developed by the loss of electrons at the initial times as reported in earlier works which are further explained in terms of double layer formation\cite{bulgakova2000double}. In nanosecond laser generated plasma, the leading portion of the laser pulse heats, melts, and vaporizes the target forming a vapor cloud. The trailing part of the laser pulse is absorbed by this vapor cloud which is mainly by inverse bremsstrahlung (IB) process. When the laser energy increases, the atoms or ions are further ionized due to electron impact ionization leading to the generation of multiple charge states. As the electrons gain large kinetic energy from the laser, they escape from the target. The loss of these highly energetic electrons leads to charge separation and creates self-generated electric field (SGEF) leading to quasi neutrality violation. These ions which get accelerated from this electric field and acquire higher kinetic energy are believed to be responsible for the fast peak observed from the LP. The second peak which is broad and extends beyond 1 $\mu$s corresponds to ions originating from slow thermal process.
Fig.~$\ref{fig:fig2}$ also shows that the ion saturation current increases with increase in laser intensities which is in line with earlier reported measurements from ion saturation current\cite{batool2021time}.\par
Interestingly, fig. $\ref{fig:fig2}$ shows a unique observation that the fast peak appears to be delayed as the laser intensity increases whereas, the slower peak becomes faster. For clarity, a pink dotted line marks the variation of peaking time of the fast and slow peaks of ion saturation current. As mentioned earlier, previous studies using Faraday cup \cite{batool2021time,bhatti2021energy} and LP \cite{irimiciuc2021langmuir,kumar2009parametric} reported that the fast and slow peaks of ion current get faster with laser intensity \cite{batool2021time,bhatti2021energy} and finally become stagnant. \cite{wu2020dynamic}. On the contrary, in our case the fast ions slow down as the laser intensity increases.\par
 \begin{figure}[ht]
\includegraphics[scale=0.92]{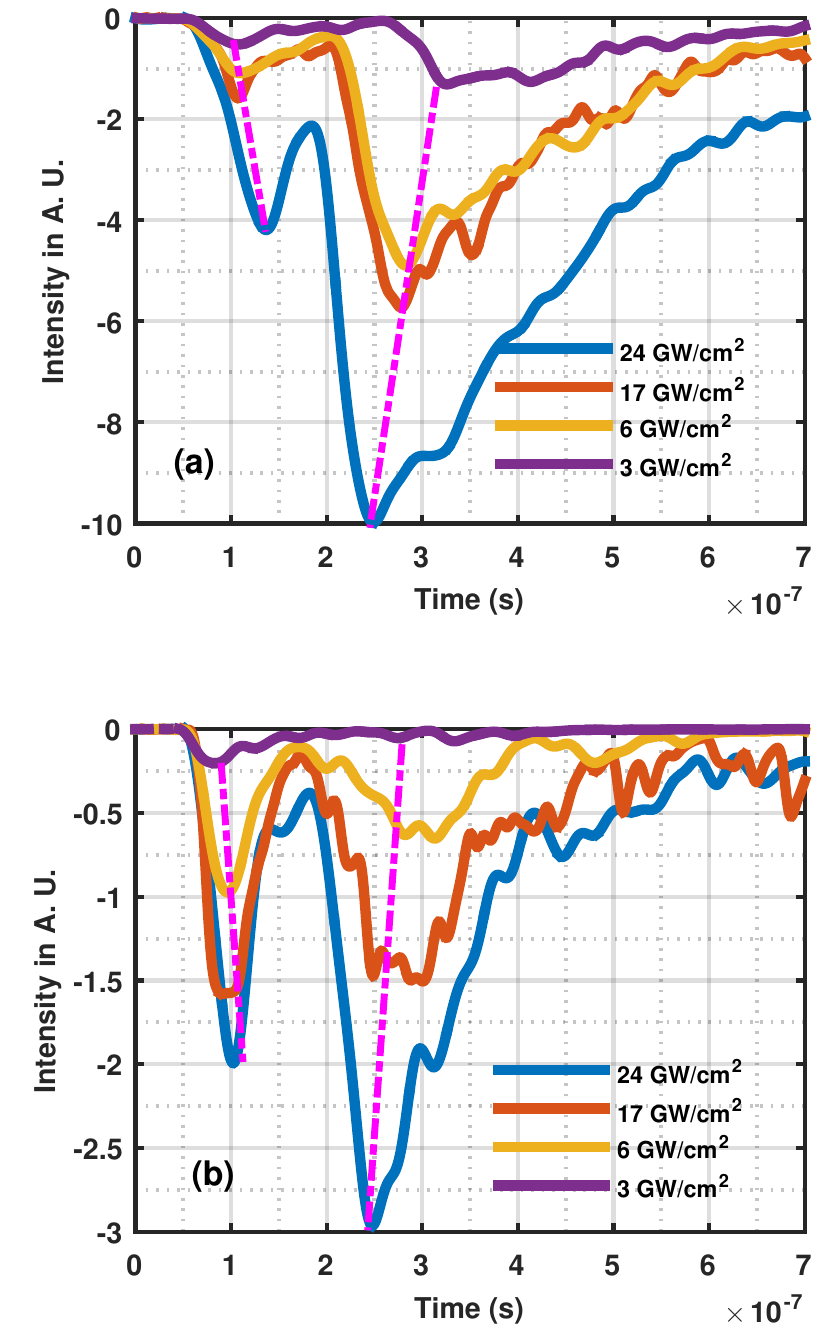}
\caption{\label{fig:fig3} STOF profiles of (a) $Al^{1+}$ (358.7 nm) ionic line  and (b) $Al^{2+}$ (451.3 nm) ionic line for four different laser intensity. The argon pressure is 0.02 mbar. The profiles are recorded at 30 mm from the target.  }
\end{figure}
To understand the unique behavior of ion saturation current with laser intensity, STOF measurements are performed for different ionic species at the same background pressure and location. Fig.~$\ref{fig:fig3}$(a) and (b) show the STOF profiles of $Al^{1+}$ and $Al^{2+}$ ionic emission for four different laser intensities. Similar to the behavior of ion saturation current, the STOF profile of these charge states also shows a double peak structure with narrow fast peak and a broad slow peak with increasing amplitude with increase in laser intensity. As can be seen from the figure, the behavior of the peaking time, (marked with pink dotted line), is similar as in case of ion saturation current (Fig.\ref{fig:fig2}). The fast peak gets delayed, whereas the slow peak becomes faster as laser intensity increases for both $Al^{1+}$ and $Al^{2+}$. This shows that the observed dynamics of ion saturation current indeed holds with individual charge species as well. \\ 
 \begin{figure}[ht]
\includegraphics[scale=1]{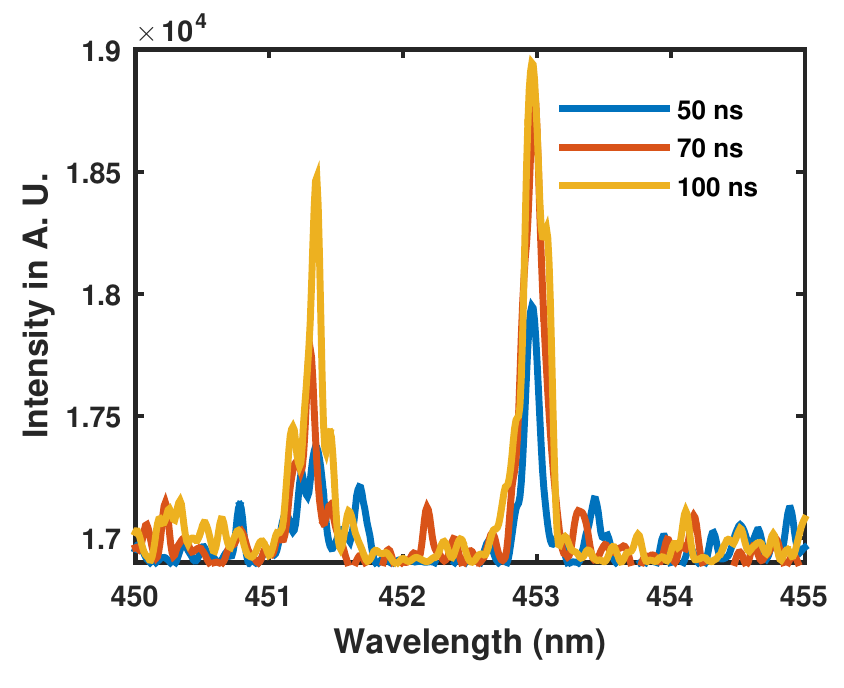}
\caption{\label{fig:fig10} Emission spectra of $Al^{2+}$ ions recorded at 30 mm from the sample at 0.02 mbar argon atmospheres for first various delays. The integration time is set to 50 ns and the plume is formed at 24 $GW/cm^2$.  }
\end{figure}
 In order to confirm that the fast peak observed in STOF is due to  ions from the target itself, the emission spectra are recorded using a spectrograph (1 meter, Mcpherson) at 30 mm distance from the sample at early times.  Fig.~$\ref{fig:fig10}$ shows the representative lines of $Al^{2+}$ recorded at different delay times within the time range of the  first peak of STOF evolution. This confirms that the the STOF spectra recorded from different charge states, indeed originate from the ionic species of the target material itself.\par
 \begin{figure}[ht]
\includegraphics[scale=0.9]{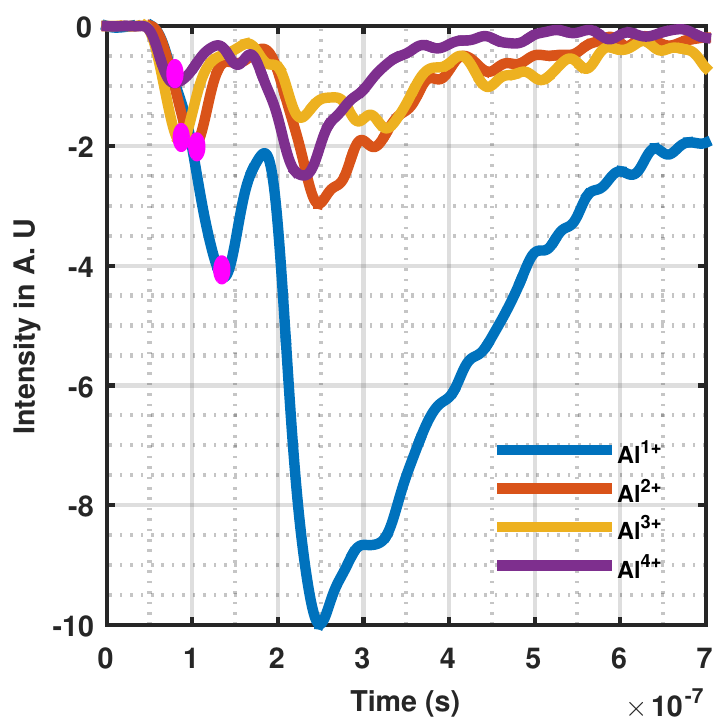}
\caption{\label{fig:fig4} STOF ionic profile of  $Al^{1+}$ (358.7 nm), $Al^{2+}$ (451.3 nm), $Al^{3+}$ (351.7 nm), $Al^{4+}$ (338.1 nm) where the plume is formed by setting laser intensity 24 GW/$cm^2$ at background pressure of 0.02 mbar . The profiles are recorded at 30 mm from the target.  }
\end{figure}
 \begin{figure}[ht]
\includegraphics[scale=0.5]{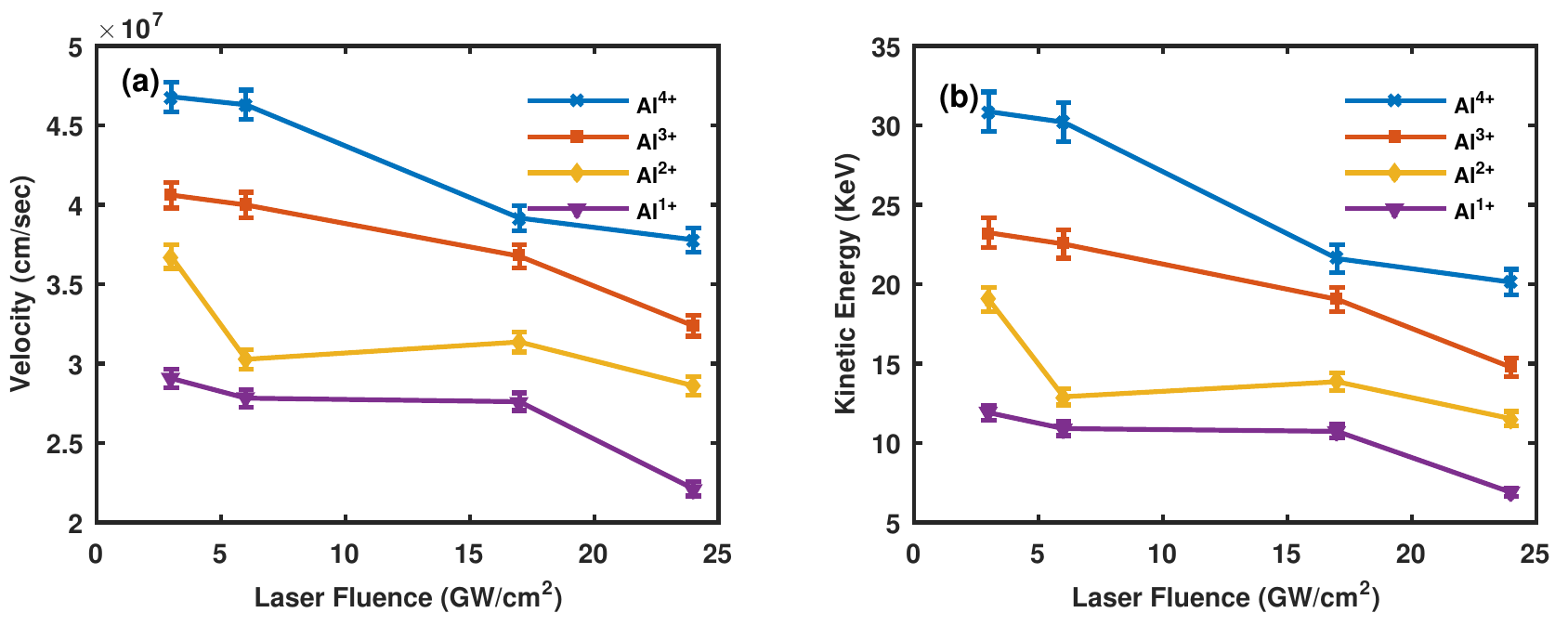}
\caption{\label{fig:fig5} (a) Velocity (b)  Kinetic energy of fast ions of $Al^{n+}$ charged species as a function of laser intensity estimated from the fast peak of STOF profile of $Al^{n+}$ ionic lines. The background pressure is set to 0.02 mbar and STOF profiles are recorded at a distance 30 mm away from the target.    }
\end{figure}
To further investigate the plasma evolution, the temporal behavior of different charge states of aluminum species produced for laser intensity of 24 GW/$cm^2$ and background $\sim$0.02 mbar is compared in Fig.\ref{fig:fig4}. It is observed that both the fast and slow components of the higher charge state reach earlier as compared to respective components of lower charge states indicating that the higher charge state acquires higher velocity. The increased velocity for higher charge states has been reported earlier and explained using the double layer \cite{wu2020dynamic,bulgakova2000double} concept where the highly charged species are expected to have larger acceleration and hence attain higher velocities. \par

Fig.~$\ref{fig:fig5}$ (a) and (b) show the trend of velocities and kinetic energies of different charged species as a function of laser intensity. The velocity of charged species is estimated from the peaking time of respective STOF profiles and subsequently kinetic energies are estimated. 
As can be seen from figure \ref{fig:fig5}, the velocity of each charged species decreases with an increase in laser intensity. However, for a particular laser intensity, the velocities and hence kinetic energies for higher charge states ions are higher than that for lower charge state ions. This again indicates that the dynamics of charged states should be governed by the electric field in the early stages of the plasma formation. It is to be noted that in our case, the energy of the fast ions of Al ions recorded at 30 mm is 10-30 keV which is significantly higher as compared to those reported earlier \cite{elsied2016characteristics,dogar2017characterization,elsied2016dynamics}. However, Javed \textit{et. al.} \cite{javed2021evaluation} reported  20-160 keV in case of fast Zr ions at 10 mm from the target for similar laser intensity range. \par
\begin{figure}[ht]
\includegraphics[scale=0.35]{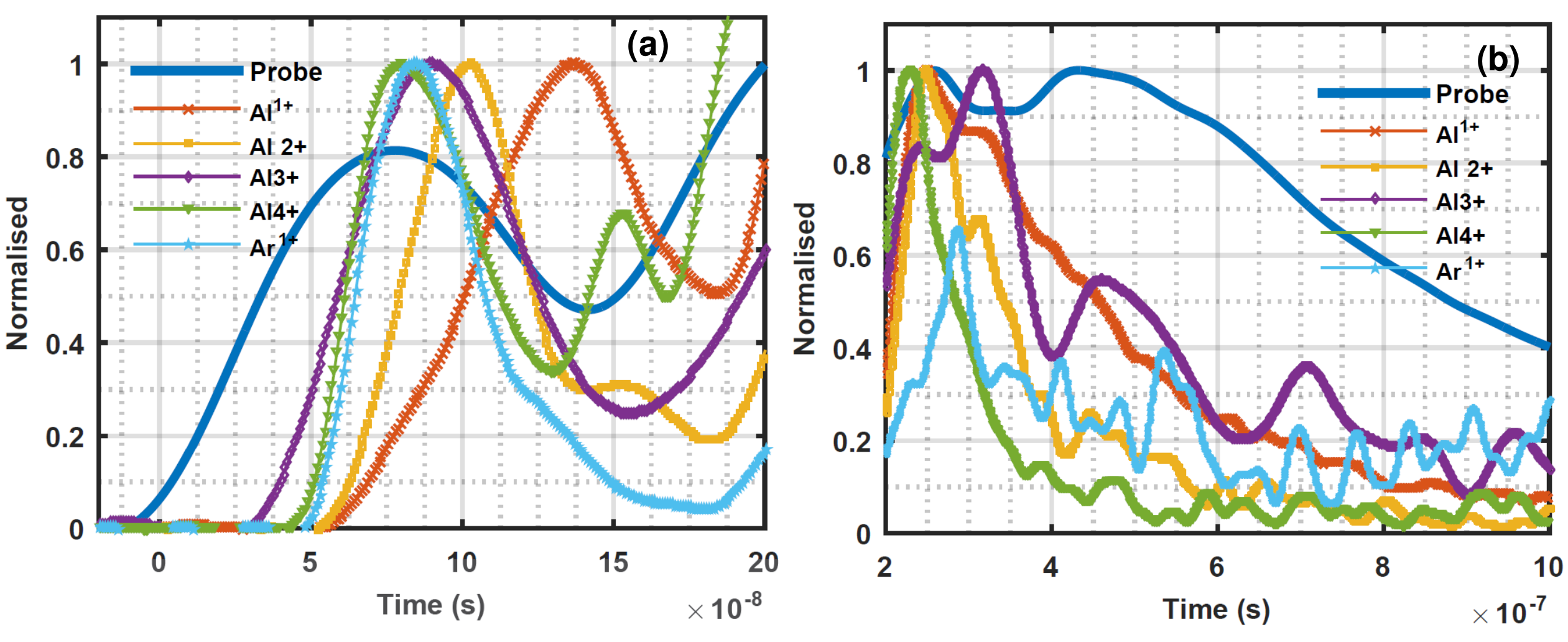}
\caption{\label{fig:fig6} Temporal evolution of  normalized ion saturation current from LP 
 as well as the normalized STOF from ionic species $Al^{1+}$,$Al^{2+}$,$Al^{3+}$,$Al^{4+}$   $Ar^{1+}$  to show the contribution of each charged species towards the ion current upto 200 ns (a) and from 200 ns to 1 $\mu$s (b). The plume is formed with 24 GW/$cm^2$ laser intensity with ambient argon pressure of 0.02 mbar. Ion current as well as STOF are recorded at 30 mm from the target.   }
\end{figure}

The ion saturation current recorded by LP is expected to have contribution from all the charged states of ions. Fig.~$\ref{fig:fig6}$ (a) (up to 200 ns) and Fig.~$\ref{fig:fig6}$(b) (from 200 ns to 2 $\mu$s) show the evolution of fast and slow peaks of ion saturation current and the STOF profiles of individual charge states of aluminum and argon respectively. The plots are normalized to the peak values of the respective species for convenient comparison. The solid blue line is the probe current, whereas the STOF emission profiles from $Al^{1+}$,$Al^{2+}$, $Al^{3+}$, $Al^{4+}$, $Ar^{1+}$ are plotted with markers joined with lines. One can notice that all the charged species contribute to the fast peak (shown in Fig.\ref{fig:fig6}(a)). The contribution to the ion saturation current is expected to be dominated by the charge states of aluminum rather than argon, considering the lower ambient pressure and dense aluminum plasma. As a result, ambient Ar is likely to be pushed outwards. However, as can be seen from figure\ref{fig:fig6}(a) $Ar^{1+}$ is present at the same time scale as that of aluminum ($Al^{4+}$) ions. The fast Ar ions can not be inherently present in the plasma plume at the same time scale as that of $Al^{1+}$ due to its heavier mass. The most probable reason for the presence of $Ar^{1+}$ in the same timescale of the fast peak of higher charge states of aluminum appears to be some kind of collisional process with highly charged states of aluminum or due to electron impact ionization. Also, it is evident from the figure that the ion saturation current is significant even at shorter time scales than the recorded evolution of charge states, which indicates that possible higher charge states (not recorded from STOF) may be present.\\
As can be seen from Fig.\ref{fig:fig6}(b) the slow peak of ion saturation current is broad and appears as a convolution of multiple peaks merged together, reaching at 3 cm around 300 ns, and persisting up to 1 $\mu$s. Here also, all the charged species are likely to contribute to the ion saturation current, and the first peak of the slow peak is precisely coinciding with the slow peaks of $Al^{1+}$ and $Al^{2+}$. The second slow peak is likely to be the contribution of all the aluminum ions and also background ions. \par
 \begin{figure}[ht]
\includegraphics[scale=0.4]{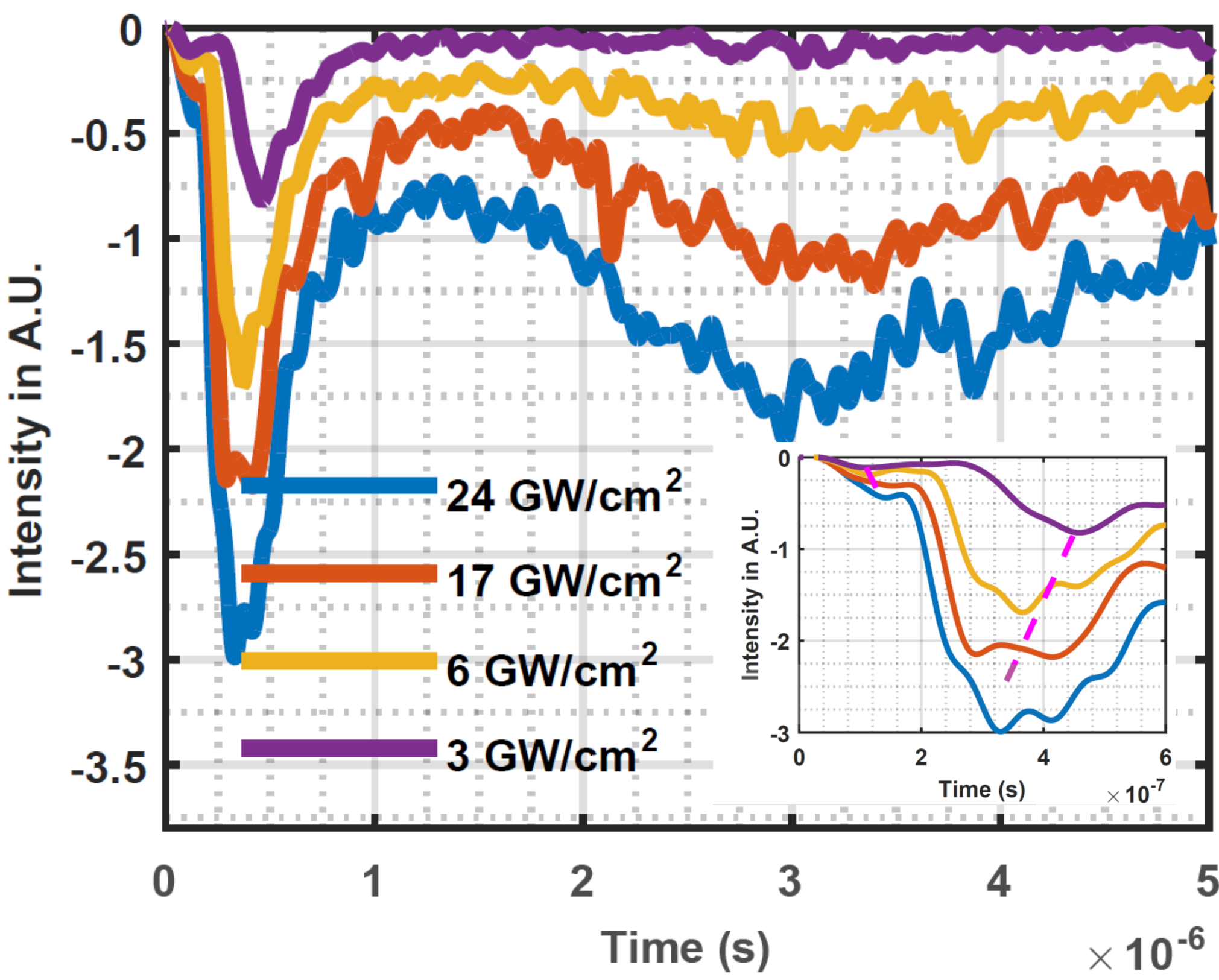}
\caption{\label{fig:fig7} STOF profile of $Al^{*}$ (396.2 nm) line for four different laser intensities. The ambient argon pressure is 0.02 mbar. The profiles are recorded at 30 mm from the target. Inset shows the STOF upto 600 ns to show the trend of fast and slow peaks with laser intensity.    }
\end{figure}
To investigate whether the observed delay in the fast peak with laser intensity is only for charged species, the time evolution of emission from $Al^{*}$ (396.2 nm) recorded for different laser intensities is shown in Fig. \ref{fig:fig7}.  It is to be noted that double peak structure is present in  the STOF of aluminum neutrals also.  Akin to aluminum ions, emission intensity increases with laser intensity.  However, as can be seen from Fig. \ref{fig:fig7}(shown in the inset), the fast peak at around 160 ns (for low intensity) and the slow peak at 250 ns again shows behavior same as that of the Al ions i.e. fast peaks slows down and the slow peaks become faster as the laser intensity increases. The fast and slow neutrals are likely to be formed by the recombination from fast and slow ions of aluminum ions respectively, and hence they appear to follow a similar trend as exhibited by aluminum ions with intensity.  Also it can be seen from the figure that STOF for neutral shows a relatively broad peak at around 3 $\mu$s. However, no emission from ions or significant ion saturation current is observed in this time range. As mentioned, the emission from aluminum neutrals is likely to originate from the recombination from Al ions as the plasma temperature decreases with time. Further, the peaking time of the broad peak is delayed as the laser intensity increases. This is because at initial stages the plasma temperature is higher at higher laser intensity, and the plume lifetime is longer. Here more ionization is expected resulting in increased charge states. However, the recombination process can occur at longer times when temperature is decreased substantially \citep{article_garima,harilalrecent}.
  \\
 \begin{figure}[ht]
\includegraphics[scale=0.4]{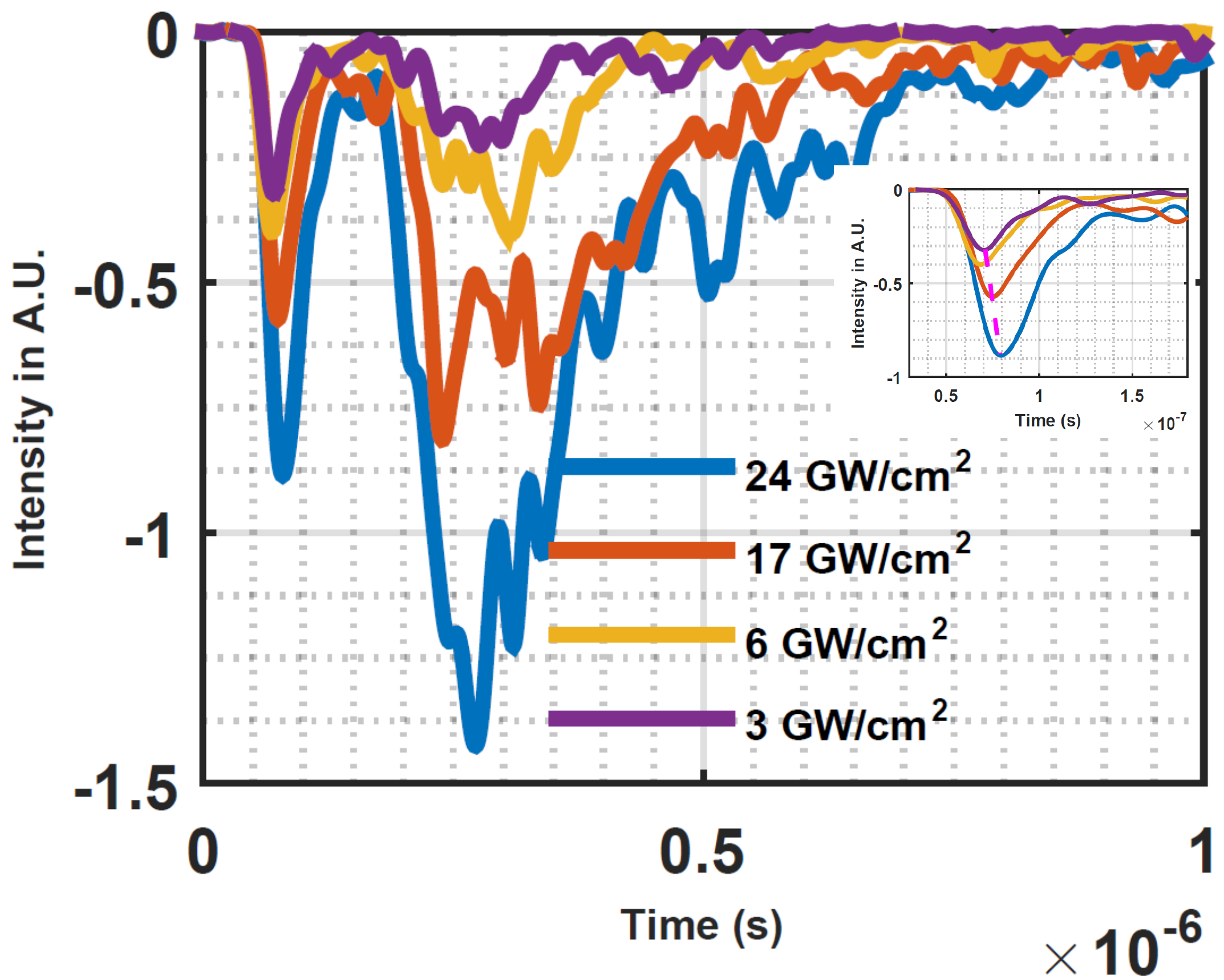}
\caption{\label{fig:fig8} STOF profile of $Ar^{*}$ (420.1 nm) line for four different laser intensities. The argon pressure 0.02 mbar. The profiles are recorded at 30 mm from the target. Inset shows the STOF upto 800 ns to demonstrate the trend of fast and slow peaks with laser intensity.     }
\end{figure}
 \begin{figure}[ht]
\includegraphics[scale=0.8]{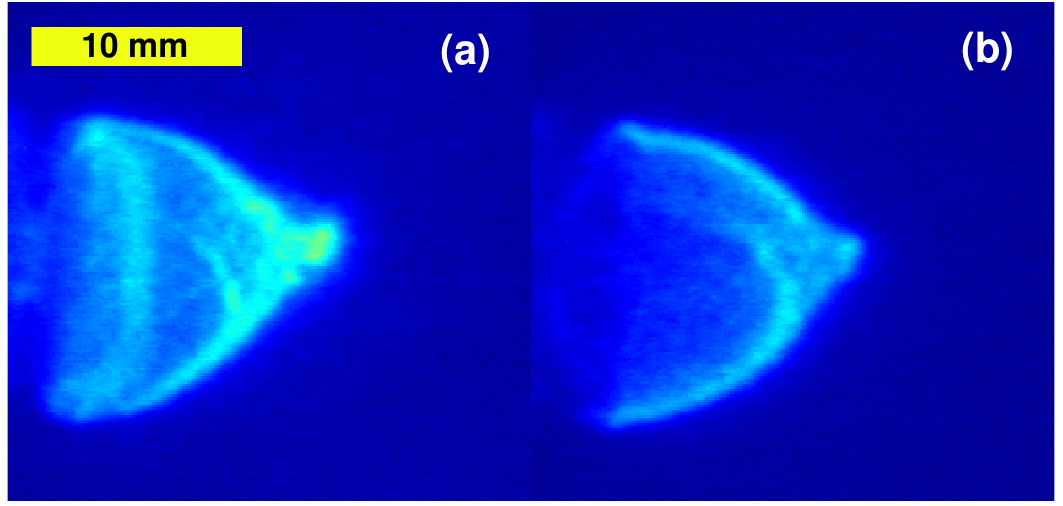}
\caption{\label{fig:fig9} ICCD images of (a) $Al^{1+}$ (466.3 nm) (b) $Ar^{1+}$ (434.8 nm) with a delay of 100 ns for an integration time of 5 ns  at 0.5 mbar argon pressure. The plume is formed with 24 GW/$cm^2$ laser power density. The filter transmission spectral range is 0.2 nm with 60$\%$ maximum transmission.   }
\end{figure}
 The temporal evolution of fast and slow peaks of Ar neutrals with an increase in laser intensity are also shown in Fig. \ref{fig:fig8}. One can notice that the emission from background argon neutrals persists upto 700 ns in contrast to Al neutrals which emits up to 6 $\mu$s. It is interesting to see that the STOF profiles of Ar neutrals also show similar trend as that of Al ions. The fast peak from Ar neutral appears around 90 ns where Al ions with higher charge states are present, and hence it is expected that these neutral may be excited by some kind of collisional process with highly charged aluminum ions although excitation by fast electrons can not be ruled out. Moreover, the ions show delay in peaking time with intensity and interestingly the fast peak of Ar neutrals appears to follow the same trend.\par
To find out the possibility of the ambient Ar penetrating the plasma plume, we have recorded the images using interference filters of $Al^{1+}$ and $Ar^{1+}$. As the emission from Argon is weak at the background pressure of 0.02 mbar, both these images are recorded at higher background pressure of 0.5 mbar.  Fig. \ref{fig:fig9} (a) and (b) show the ICCD images of aluminum and argon ions at a time delay of 100 ns, respectively. It is clearly observed from the images that the contribution of emission in case of $Al^{1+}$ is present in the main plasma plume and its boundary (near to the shock region). However, in case of  $Ar^{1+}$ the emission intensity is confined only to the boundary (periphery) of the plume even at a higher background pressures. This further supports the fact that the background ions do not constitute the main plasma plume and are present at the boundary. Further, these are excited by the electron or ion impact processes.\par
Briefly, from the present work, it is clearly evident that the velocity of fast ions decreases with increase in laser intensity. Here, it is anticipated that the faster peak due to self generated electric fields at initial stage of plasma formation should become more prominent for higher laser intensities and there should be increase in the velocity of the charged species depending upon the charge. Interestingly, an opposite trend is observed in the present experiments. Here we would like to mention that in an earlier work, Wu \textit{et. al.}\cite{wu2020dynamic} found that there is a decrease in the kinetic energy of Mo$^{+1}$ ions as the laser intensity increases, which, however, it is observed for higher charge states in our work. They tentatively suggested that this behavior is likely to occur due to more collisions and recombinations during the expansion process. They also attributed the decrease in kinetic energy with intensity due to plasma shielding and absorption. In fact, the present experiments also have comparable laser intensity. From a simple comparison, a similar reason can be expected for this observed reduction in kinetic energy of aluminum charge states. However, we would like to mention that in our case we have observed decrease in energy for all the studied charged states. Hence, it can not be pointed out if there is any substantial evidence of the role of recombination processes.\\
Another possibility for the observed decrease in velocity is the decrease in the field strength with increase in laser inetnsity at the early stages of plasma formation. It can be expected that the initial plasma plume density in the self regulating regime \cite{harilal1997electron} may restrict the maximum plasma density to the critical density which depends on the laser frequency irrespective of the laser intensity. However, at higher laser intensities, the temperature may increase due to the larger IB process. Hence, it can be anticipated that during the interaction of laser pulse with the sample, the plasma density remains the same but the temperature  increases  as the laser intensity increases. In this scenario, Debye length of plasma plume will be larger for the plasma formed with higher laser intensity. As the thickness of the double layer formation depends on the Debye length, there is a possibility that the electric field strength may be weakened with higher laser intensity, and subsequent decrease in the velocity of charged species is expected. It can be mentioned that at higher peak intensity, the rising edge of the laser pulse becomes high enough to result in the plasma expansion and hence form the electric field. However, later the  intensity does not appear to contribute to charge separation probably due to decreased plasma density or spatial mismatch between the plasma and the laser focal spot. As a result, the temporal structure of the laser pulse and evolution of the laser-plasma interaction can be important.\\
Third possibility is the presence of bipolar nature of SGEF with opposite polarities as reported in some of the recent studies \cite{batool2021time,bhatti2021energy}. With an increase in intensity, the balance between these fields with opposite polarities is likely to be disturbed and hence a decrease in the velocities of charged species is possible. Here it can be mentioned that for correctly ascertaining the observed behavior of the velocity of charged species, more insights into the plasma parameters and hence electric fields at the very early stage of plasma are needed. As the electric field can only be present within the order of few Debye lengths, the density and temperature of plasma at the very initial stages of plume expansion is crucial. However, the existing diagnostics are not capable to get the plasma parameters at the very early stages of expansion. Though a clear cut explanation for the observed behavior can not be given at present, we believe the present work points towards a peculiar behavior of fast ions which may lead to further investigations in this direction.\\
Although it is not possible to exactly arrive at the exact mechanism of the delay in the fast peak at this stage, the following possibilities can be argued. As it was not possible to the experiments in ultra high vacuum, we can not rule out if background gas has a role in it. However, delay in the fast peak is observed for higher background pressures. Second possibility may arise due to the creation of electric field at the rising edge of the laser pulse, which somehow appears not contributing to charge separation at higher intensities. This may happen because of decreased plasma density or spatial mismatch between plasma and laser focal spot.    
\section{Conclusions}\label{sec:conc} 
The dynamics of fast and slow ions is studied for different laser intensities using LP and STOF. The ion current from LP shows a double peak structure where the first peak is sharp and peaks around 100 ns. However, the slow peak is broad with peaking time around 400 ns. The ion current increases with laser intensity and the slow peak advances in time for higher laser intensity, similar to earlier reported works using ion collector current from FC. However in the present study, the fast peak shows somewhat peculiar behavior with laser intensity i.e it gets delayed in time with increase in laser intensity. Further, ion saturation current which can be assumed as the convolution of the contributions from all the charged species and confirms the anomaly in the fast peak. STOF profiles of all the charged species show similar trend as observed in case of LP. The double peak nature is present in the case of higher charged species in which fast peak is narrow and slow peak is broad. The decrease in velocity and hence kinetic energy is also observed in STOF as in case of LP. The energy of fast ions is the range of 10-40 KeV. The higher charged species have higher velocities as compared to lower charged species, which confirms the presence of  transient electric fields. Another interesting aspect is the dynamics of the background argon in the expanding plume. Though more detailed investigations of the early stages of plume expansion are required for getting finer aspects of the mechanism, we believe the present study brings out some hitherto unexplored interesting features regarding the species present in laser produced aluminum plasma plume in the presence of ambient argon.   

\section*{Data availability} 

The data that supports the observations of this study are available from the corresponding author upon reasonable request.

\section{references}
%

\end{document}